\documentclass[reprint,amsmath,amssymb,aps,superscriptaddress,pra]{revtex4-1}

\usepackage[T1]{fontenc}

\usepackage[english]{babel}
\usepackage[utf8]{inputenc}
\usepackage{amsmath}
\usepackage{graphicx}
\usepackage[colorinlistoftodos]{todonotes}

\usepackage{hyperref}
\usepackage{bm}
\usepackage{ulem}
\usepackage{comment}

\newcommand{\p}{\partial}
\newcommand{\half}{\frac{1}{2}}

\newcommand{\e}{\bm{\hat{e}}}

\newcommand{\ez}{\bm{\hat{e}}_3}
\newcommand{\emu}{\bm{\hat{e}}_\mu}
\newcommand{\emn}{\epsilon_{\mu\nu}}

\newcommand{\magn}{\bm{m}}

\renewcommand{\v}{\tilde{v}}
\newcommand{\s}{\tilde{s}}

\newcommand{\Exchange}{A}
\newcommand{\DM}{D}
\newcommand{\dm}{\lambda}

\newcommand{\Anisotropy}{K}
\newcommand{\anisotropy}{\kappa}

\newcommand{\Energy}{E}
\newcommand{\Eex}{\Energy_{\rm ex}}
\newcommand{\Ean}{\Energy_{\rm a}}
\newcommand{\Edm}{\Energy_{\rm DM}}
\newcommand{\Edmm}{\tilde{\Energy}_{\rm DM}}
\newcommand{\Hext}{H}
\newcommand{\hext}{h}
\newcommand{\bHext}{\mathbf{H}}
\newcommand{\bhext}{\mathbf{h}}
\newcommand{\hmagn}{\mathbf{h}_m}

\newcommand{\lex}{\ell_{\rm ex}}
\newcommand{\ldm}{\ell_{\rm D}}
\newcommand{\ldw}{\ell_{\rm w}}
\newcommand{\ldk}{\ell_{S}}
\newcommand{\thickness}{d_f}

\begin{document}

\preprint{APS/123-QED}
 
\title{Chiral magnetic skyrmions across length scales}
\author{Stavros Komineas}
\affiliation{Department of Mathematics and Applied Mathematics, University of Crete, 70013 Heraklion, Crete, Greece}
\author{Christof Melcher}
\affiliation{Department of Mathematics I \& JARA Fundamentals of Future Information Technology, RWTH Aachen University, 52056 Aachen, Germany}
\author{Stephanos Venakides}
\affiliation{Department of Mathematics, Duke University, Durham, NC, USA}
\date{\today}

\begin{abstract}
The profile and energy of chiral skyrmions, found in magnetic materials with the Dzyaloshinskii-Moriya interaction, can be approximated by formulae obtained through asymptotic analysis in the limits of small and large skyrmion radius.
Using fitting techniques, we modify these formulae so that their validity extends to almost the entire range of skyrmion radii.
Such formulae are obtained for skyrmions that are stabilised in the presence of an external field or easy-axis anisotropy or a combination of these.
We further study the effect of the magnetostatic field on the skyrmion profile.
We compare the profile of magnetic bubbles, stabilized without the chiral Dzyaloshinskii-Moriya interaction to that of a chiral skyrmion.
\end{abstract}


\maketitle

\section{Introduction}
\label{sec:Intro}

Chiral magnetic skyrmions are topological magnetic configurations that are stabilised in materials with the Dzyaloshinskii-Moriya interaction \cite{EverschorMasellReeveKlaeui_JAP2018}.
They have been observed in a number of experiments and the detailed features of individual skyrmions have been resolved experimentally to an impressive degree for isolated skyrmions \cite{2015_PRL_RommingKubetzkaWiesendanger,BoulleVogel_nnano2016,2016_NJP_LeonovBogdanovWiesendanger,KovacsBorkovski_PRL2017,ShibataTokura_PRL2017,2019_NatComm_MeyerPerini} and in a skyrmion lattice \cite{2016_NJP_McGroutherLamb}.
The experimental works have mapped the profile of the skyrmion, i.e., the magnetization as a function of the distance from its center.
The skyrmion profile is the foundation for the study of the statics and of dynamical behaviors of skyrmions and the subsequent derivation of quantitative results.
The profile enters in formulae for dynamical phenomena, for example, skyrmion translation and oscillation modes \cite{2018_PRB_KravchukShekaGaididei,SchuetteGarst_PRB2014} or antiferromagnetic skyrmion excitations \cite{2019_PRB_KravchukGomonaySheka}, and it is crucial for quantitative calculations.
Furthermore, it quantifies the localization of the entity which in turn establishes its particle-like nature.

Approximation formulae that capture basic features of the skyrmion profile have been proposed.
For skyrmions of large radius, an ad-hoc ansatz based on explicit one-dimensional domain wall profiles \cite{Braun_PRB1994} has been suggested and is widely used to examine structural and dynamic properties \cite{2013_PRB_RohartThiaville,2015_PRL_RommingKubetzkaWiesendanger,2015_NatComm_Zhou,2018_SciRep_BuettnerLemeshBeach}.
The domain wall profile with an adjustable wall thickness  \cite{2018_CommPhys_WangYuan,2018_PRB_KravchukShekaGaididei} and a related ansatz \cite{2016_NJP_LeonovBogdanovWiesendanger} have been employed as more flexible forms in order to capture the skyrmion features over a wider parameter range.
For the case of a model with easy-axis anisotropy, accurate skyrmion profiles in the asymptotic sense have been obtained for skyrmions of small \cite{2020_NL_KomineasMelcherVenakides} and of large radius \cite{2021_PhysD_KomineasMelcherVenakides}.
Similar methods were used for a model including the magnetostatic field \cite{2020_ARMA_BernandMuratovSimon,2020_PRB_BernandMuratovSimon}.
In the present paper, we give a comprehensive analysis of the skyrmion profile for the case of easy-axis anisotropy and external magnetic field.
We use the methods that were developed for the case of easy-axis anisotropy only in Refs.~\cite{2020_NL_KomineasMelcherVenakides,2021_PhysD_KomineasMelcherVenakides}.
While the work in the above references focuses on asymptotic formulae for skyrmions of small and large radius, the present paper gives a set of formulae for determining the skyrmion radius that are good approximations in the whole parameter range, i.e., also for intermediate radii.
The approximation formulae are derived so as to be consistent with the asymptotic ones.
We further present the result of a set of numerical simulations for the effect of the magnetostatic field on the axially-symmetric skyrmion.
We make a direct comparison between the profiles of chiral skyrmions and bubbles which reveals that, beyond basic differences, the two entities share many features.

We, finally, make a detailed study of the skyrmion energy, made possible by the availability of a detailed analytical description of the skyrmion profile.
The analytical calculations and formulae will help focus the efforts for applications of skyrmions.

The paper is arranged as follows.
Sec.~\ref{sec:formulation} formulates the problem.
Sec.~\ref{sec:farField} gives the far field for the skyrmion profile.
Sec.~\ref{sec:externalField} gives the formulae for the skyrmion profile and radius for the case of an external field.
Sec.~\ref{sec:anisotropy} gives the corresponding formulae for the case of easy-axis anisotropy.
Sec.~\ref{sec:magnetostatic} studies the effect of the magnetostatic field on the skyrmion and gives comparisons with magnetic bubbles.
Sec.~\ref{sec:conclusions} contains our concluding remarks.
In the Appendix, we give a complete review of the method of asymptotic matching and we expand it so that it can readily be applied for the derivation of formulae in the main text.

\section{Formulation}
\label{sec:formulation}

We consider a thin film of a ferromagnetic material on the $xy$-plane with symmetric exchange, Dzyaloshinskii-Moriya (DM) interaction, anisotropy of the easy-axis type perpendicular to the film, and an external field.
The micromagnetic structure is described via the magnetization vector $\magn=\magn(x,y)$ with a fixed length normalized to unity, $\magn^2=1$.
The energy is
\begin{equation} \label{eq:energy0}
\begin{split}
    \Energy = & \Exchange \int (\p_\mu\magn\cdot\p_\mu\magn) d^2 x + \DM \int e_{\rm DM}\,d^2x \\
    & + \Anisotropy \int (1-m_3^2) d^2x - \mu_0 M_s^2 \int \bhext\cdot\magn\,d^2x
\end{split}
\end{equation}
where $\Exchange, \DM, \Anisotropy$ are the exchange, DM, and anisotropy parameters, respectively, and the external field $\bhext=\bHext/M_s$ is normalized to the saturation magnetization $M_s$.
The DM energy density $e_{\rm DM}$ may be chosen to have the bulk form $e_{\rm DM }= \emu\cdot(\p_\mu\magn\times\magn)$ or the interfacial form $e_{\rm DM} = \emn \emu\cdot(\p_\nu\magn\times\magn)$, where $\mu,\nu=1,2$ and the summation convention is invoked, while $\emu$ are the unit vectors for the magnetization in the respective directions.

It is useful to consider the length scales that arise naturally,
\begin{equation} \label{eq:lengthScales}
    \lex = \sqrt{\frac{2A}{\mu_0 M_s^2}},\quad
    \ldw = \sqrt{\frac{A}{\Anisotropy}},\quad
    \ldm = \frac{2A}{\DM},\quad
    \ldk = \frac{\DM}{2\Anisotropy}.
\end{equation}
$\lex$ is the exchange length, $\ldw$ is the domain wall width, $\ldm$ gives the pitch of the spiral solution for strong DM interaction, and $\ldk$ is related to the radius of small skyrmions.
Using $\lex$ as the unit of length, and assuming an external field $\bhext=\hext\ez$ perpendicular to the film, the scaled form of the energy is
\begin{equation} \label{eq:energy}
\begin{split}
    \Energy = & \half \int (\p_\mu\magn\cdot\p_\mu\magn) d^2 x + \dm \int e_{\rm DM}\,d^2x \\
    & + \frac{\anisotropy}{2} \int (1-m_3^2) d^2x + \hext \int (1-m_3)\,d^2x
\end{split}
\end{equation}
where we have assumed that $\magn=\ez$ in the uniform magnetization of the film and we have used the scaled DM and anisotropy parameters
\begin{equation} \label{eq:parameters}
    \dm = \frac{\lex}{\ldm} = \frac{\DM}{\sqrt{2\Exchange\mu_0 M_s^2}},\qquad
    \anisotropy = \frac{\lex^2}{\ldw^2} = \frac{2\Anisotropy}{\mu_0 M_s^2}.
\end{equation}
Static magnetization configurations satisfy the time-independent Landau-Lifshitz equation
\begin{equation} \label{eq:LL}
\magn \times \left( \p_\mu\p_\mu\magn - 2\dm\, \mathbf{h}_{\rm DM} + \anisotropy\,m_3 \ez + \hext\ez \right) = 0.
\end{equation}
The bulk DM form is $\mathbf{h}_{\rm DM} = \emu\times\p_\mu\magn$ and the interfacial one is $\mathbf{h}_{\rm DM} = \emn \emu\times\p_\nu\magn$.


Skyrmion solutions with axial symmetry can be described in polar coordinates $(r,\phi)$ using the spherical angles ($\Theta, \Phi)$ for the magnetization.
We choose $\Phi=\phi$ for the case of interfacial DM interaction and $\Phi=\phi+\pi/2$ for the case of bulk DM interaction.
In both cases, the angle $\Theta=\Theta(r)$ satisfies
\begin{equation} \label{eq:thetaODE}
    \Theta'' + \frac{\Theta'}{r} - \frac{\sin(2\Theta)}{2 r^2} + 2\dm \frac{\sin^2\Theta}{r}- \frac{\anisotropy}{2} \sin(2\Theta) - \hext \sin\Theta = 0.
\end{equation}
Existence, uniqueness, stability and minimality of axially symmetric skyrmions in the regime $\lambda^2/ \kappa\ll 1$ and
$\lambda^2/h \ll 1$, respectively, has been examined rigorously in \cite{2018_LiMelcher, 2022_GustafsonWang}.
In the following, we will use the convention that $\Theta(r=0)=\pi$ in the skyrmion center and $\Theta = 0$ at spatial infinity $r\to\infty$.
The skyrmion radius $R$ will be defined to be at the radial distance where $\Theta=\pi/2$.

\section{Far-field}
\label{sec:farField}

In the far field, where $\Theta$ is small, Eq.~\eqref{eq:thetaODE} reduces by linearization to
\begin{equation} \label{eq:modifiedBessel_theta}
r^2\Theta''+r\Theta' - [1 + (\anisotropy+\hext) r^2]\Theta = 0.
\end{equation}
Under a scaling transformation $\sqrt{\anisotropy+\hext}\,r \to r$ this gives the modified Bessel equation.
The appropriate solution, decaying for $r\to\infty$, is the modified Bessel function of the second kind $K_1(r)$ \cite{AbramowitzStegun}, that also appeared in a similar context in Ref.~\cite{VoronovIvanovKosevich_JETP1983}.
We have
\begin{equation} \label{eq:K1_theta}
\begin{split}
\Theta(r) = & C\,K_1(r) = \frac{C}{r} \left[ 1 + (2\gamma-1) \left( \frac{r}{2} \right)^2  \right. \\
  & \left. + 2 \sum_{n=1}^\infty \frac{(\ln\left( \frac{r}{2} \right) - \ln n) \left( \frac{r}{2} \right)^{2n}}{(n-1)!\, n!} - 2 \sum_{n=2}^\infty \frac{\xi_n \left( \frac{r}{2} \right)^{2n}}{(n-1)!\, n!} \right]
\end{split}
\end{equation}
where $C$ is an arbitrary constant, $\gamma \approx 0.57721$ is the Euler-Mascheroni constant, and
\begin{equation} \label{eq:Euler}
\begin{split}
\xi_n & = \frac{1}{2} (\psi(n)+\psi(n+1))-\ln n, \\
\psi(n) & = -\gamma+\sum_{k=1}^{n-1}\frac{1}{k}, \quad n=2,\ldots.
\end{split}
\end{equation}

The asymptotic form of $K_1(r)$ for large values of $r$ gives \cite{AbramowitzStegun}
\begin{equation} \label{eq:exponentialDecay}
\Theta \sim C \sqrt{\frac{\pi}{2r}}\;e^{-r},\quad r\gg 1.
\end{equation}
It is interesting to note that the result for the far field does not depend on the DM interaction.
The exponential decay \eqref{eq:exponentialDecay} is due to the anisotropy and the external field.

\section{External field}
\label{sec:externalField}

We consider that we only have an external field and no anisotropy, $\anisotropy=0$.
Then, we may choose $\lex\sqrt{M_s/\Hext}$ as the unit of length which leads to setting $\hext \to 1$ in Eq.~\eqref{eq:thetaODE} and the DM parameter is
\begin{equation}  \label{eq:parameter_hext}
\dm = \frac{\lex}{\ldm} \sqrt{\frac{M_s}{\Hext}} = \frac{\DM}{\sqrt{2A\mu_0 M_s\Hext}}.
\end{equation}

We initially focus on skyrmions of small radius.
In this limit, it is convenient to work with the variable
\begin{equation} \label{eq:stereo}
    u(r) = \tan\frac{\Theta(r)}{2}
\end{equation}
which is the modulus of the stereographic projection of the magnetization vector.
In the case, of exchange interaction only, we have the well-known axially-symmetric Belavin-Polyakov (BP) solution \cite{BelavinPolyakov_JETP1975}
\begin{equation}  \label{eq:BPskyrmion}
u(r) = \frac{R}{r},
\end{equation}
that represents a skyrmion of unit degree with a radius $R$. 
Inspired by the BP solution \eqref{eq:BPskyrmion}, we remove the singularity at the origin by defining the field
\begin{equation}  \label{eq:v}
    \v(r) = r u(r).
\end{equation}
For comparison purposes, we note that the variable $v$ in Ref.~\cite{2020_NL_KomineasMelcherVenakides} is related to the present $\v$ by $\v=2v$. 

The equation for $\v$ is given in Appendix~\ref{sec:asymptotics} in Eq.~\eqref{eq:v-prime}.
In Appendix~\ref{sec:nearField}, we consider the case of small values of $\dm$ and find the near field at the center of the skyrmion, in Eqs.~\eqref{eq:nearField_wtau}, and well beyond its radius, in Eq.~\eqref{eq:nearField_v_matching}.
This generalises the results of Ref.~\cite{2020_NL_KomineasMelcherVenakides} for the case where both an external field and an anisotropy are present.
The analysis is based on the fact that $\v(r)$ is small in the entire space, when $R$ is small.
In Appendix~\ref{sec:matching}, by matching the near field of Eq.~\eqref{eq:nearField_v_matching} with the far field in Eq.~\eqref{eq:K1_theta} in a region where both are valid, a relation between the system parameters and the skyrmion radius is obtained in Eq.~\eqref{eq:dm_R}.

For $\hext=1$ and $\anisotropy=0$, Eq.~\eqref{eq:dm_R} reduces to
\begin{equation} \label{eq:dm_v0-hfield}
    \dm = -R \left[ \gamma + \frac{1}{2} + \ln(R/2) \right]
\end{equation}
or
\begin{equation} \label{eq:dmR_field}
  \dm = -R\, \ln\left( \frac{R}{\alpha_\hext} \right),
  \qquad \alpha_\hext = 2 e^{-(\gamma+\frac{1}{2})} \approx 0.6811.
\end{equation}

\begin{figure}
    \centering
    (a)\includegraphics[width=8cm]{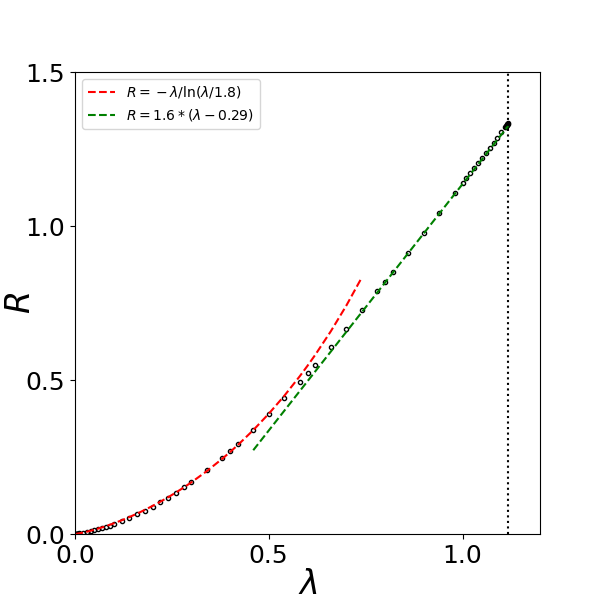}
    (b)\includegraphics[width=8cm]{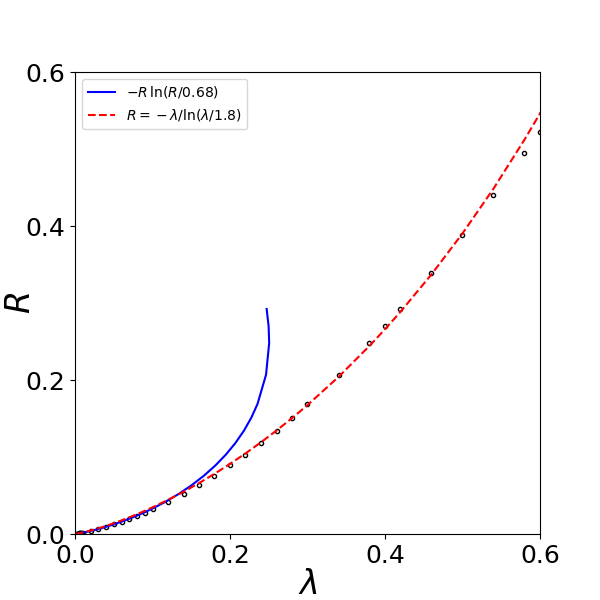}
    \caption{(a) The skyrmion radius $R$, in units of $\lex\sqrt{M_s/\Hext}$, found numerically for various values of $\dm$, defined in Eq.~\eqref{eq:parameter_hext}, is shown by open circles.
    The red line shows the fitting formula \eqref{eq:dmR_field_fit}.
    The green line shows the fitting formulae \eqref{eq:dmR_field_fit_large}.
    The dotted vertical line marks the critical value $\dm=\dm_0\approx 1.12$ beyond which no isolated skyrmions exist.
    Figure (b) is a blow up of figure (a) for small $\dm$ values.
    The blue line shows the asymptotic formula \eqref{eq:dmR_field} for small radius.
    }
    \label{fig:dm-R_field}
\end{figure}

We solve Eq.~\eqref{eq:v-prime} (or Eq.~\eqref{eq:thetaODE}) numerically using a shooting method and we find the skyrmion profiles for various values of the parameter $\dm$.
Given the profile, we detect the skyrmion radius.
In Fig.~\ref{fig:dm-R_field} we show the skyrmion radius found numerically as a function of the parameter.
Formula \eqref{eq:dmR_field} is given in entry (b) of the figure and it fits excellently the numerically found skyrmion radius for small $R$.

Isolated skyrmions exist in this model for $\dm \leq \dm_0 \approx 1.12$ while at $\dm = \dm_0$ a transition to a skyrmion lattice occurs \cite{BogdanovHubert_JMMM1994}.
Inverting relation \eqref{eq:dmR_field} gives the leading order of approximation
\begin{equation} \label{eq:dm-R_small}
    R = -\frac{\dm}{\ln\dm}.
\end{equation}
Following Ref.~\cite{2022_ZimmermannBluegel}, a modification of the denominator of Eq.~\eqref{eq:dm-R_small},
\begin{equation} \label{eq:dmR_field_fit}
    R = - \frac{\dm}{\ln(\dm/1.8)}
\end{equation}
leads to a very good fit of the numerical data over a wide range as shown in Fig.~\ref{fig:dm-R_field}.
Eq.~\eqref{eq:dmR_field_fit} is consistent with Eq.~\eqref{eq:dmR_field} at $\dm\to 0$, but the factor 1.8 is found only by fitting the numerical data.
For larger $\dm$, the linear relation
\begin{equation} \label{eq:dmR_field_fit_large}
    R = 1.6(\dm-0.29)
\end{equation}
fits the numerical data.
The combination of formulae \eqref{eq:dmR_field_fit} for $\dm \lesssim 0.6$ and \eqref{eq:dmR_field_fit_large} for $\dm \gtrsim 0.6$ give a very good fit of the skyrmion radius for the entire range of the parameter space.

We now turn to the skyrmion profiles.
Eq.~\eqref{eq:nearField_wtau} gives the profile at the skyrmion inner core and Eq.~\eqref{eq:K1_theta} gives the profile at large distances.
Based on the above asymptotic results, we will give simple approximation formulae for the skyrmion profile.

The field around the skyrmion center is found by a Taylor expansion of the solution in Eq.~\eqref{eq:nearField_wtau} close to $r=0$ (applied for $\hext=1, \anisotropy=0$) while the parameter $\dm$ is substituted from Eq.~\eqref{eq:dm_v0-hfield}.
The calculations are given in Appendix~\ref{sec:innerCore} and lead to Eq.~\eqref{eq:innerCoreField}.
For the present case, this reduces to the approximation
\begin{equation} \label{eq:core_theta_small}
    \tan\left(\frac{\Theta}{2}\right) = \frac{R}{r} \left[ 1 + R^2\,\left( \gamma + \frac{3}{4} + \ln(R/2)  \right)\, \frac{r^2}{2} \right]
\end{equation}
and it is valid for $r < R$ for small radius skyrmions.

In the far field, $\Theta$ and the field $u$ in Eq.~\eqref{eq:stereo} are proportional to each other and both satisfy the modified Bessel equation \eqref{eq:modifiedBessel_theta}.
We will continue the discussion using the field $u$, instead of $\Theta$, because it eventually gives a good fit for the skyrmion profile over a wider interval in $r$.
For small $R$, we have $u = R K_1(r)$, where the factor $R$ follows from the asymptotic matching conditions.
We write
\begin{equation} \label{eq:farField_theta_small}
    \tan\left(\frac{\Theta}{2}\right) = R\,K_1(r),\quad r \gg R.
\end{equation}

\begin{figure}
    \centering
    \includegraphics[width=8cm]{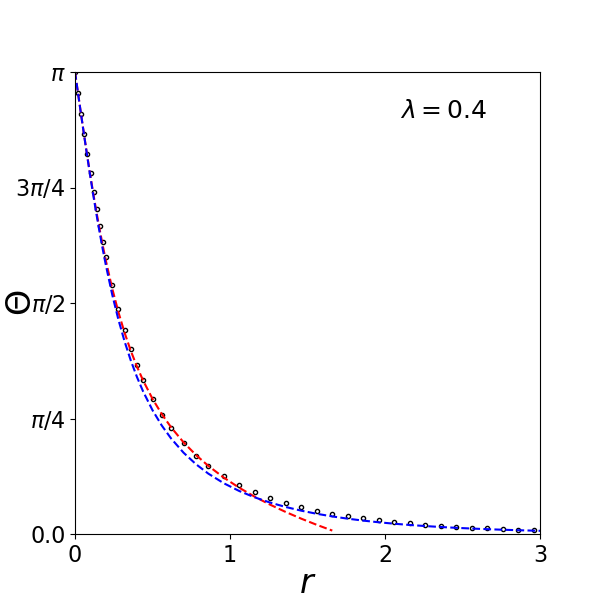}
    \caption{Numerically calculated skyrmion profiles $\Theta(r)$ are shown by open circles for the case of external field and parameter value $\dm=0.4$.
    The red dashed line shows formula \eqref{eq:core_theta_small} that is valid close to the skyrmion center and the blue dashed line shows the far field formula \eqref{eq:farField_theta_small}.
    Both formulae turn out to give good approximations far beyond their expected range of asymptotic validity.
    Lengths are measures in units of $\lex\sqrt{M_s/\Hext}$.}
    \label{fig:profiles_field_small}
\end{figure}

Fig.~\ref{fig:profiles_field_small} shows the profile for a skyrmion of small radius.
Surprisingly, the far field formula \eqref{eq:farField_theta_small}, shown by the blue dashed line, gives a good approximation almost in the entire space.
Close to $r=0$, Eq.~\eqref{eq:core_theta_small} is the correct approximation.
Eq.~\eqref{eq:farField_theta_small} cannot be justified in this region.

We conclude the section by giving a description of the complete skyrmion profile over the full spatial range.
A validity condition for the near field is $\dm r^2 \ll \v_0$ \cite{2020_NL_KomineasMelcherVenakides}.
Using Eq.~\eqref{eq:dmR_field} and $\v_0\approx R$, we have that Eq.~\eqref{eq:core_theta_small} is valid for 
\begin{equation} \label{eq:BPcondition}
    r^2\ll -\frac{1}{\ln R}.
\end{equation}
The form of Eq.~\eqref{eq:core_theta_small} indicates that in the regime where condition \eqref{eq:BPcondition} is valid the skyrmion profile is close to the BP profile.
The exponentially decaying behaviour \eqref{eq:exponentialDecay} is valid for $r\gg 1$.
Between the regimes of validity of the BP profile and the exponentially decaying profile, the modified Bessel function \eqref{eq:farField_theta_small} is still a good approximation.

\section{Easy-axis anisotropy}
\label{sec:anisotropy}

We consider the case of easy-axis anisotropy and no external field, $\hext=0$.
Then, we may choose $\ldw$ as the unit of length which leads to setting $\anisotropy \to 1$ in Eq.~\eqref{eq:thetaODE} and
\begin{equation}  \label{eq:parameter_aniso}
\dm = \frac{\ldk}{\ldw} = \frac{\DM}{2\sqrt{A\Anisotropy}}.
\end{equation}
This section is based on the work presented in Refs.~\cite{2020_NL_KomineasMelcherVenakides,2021_PhysD_KomineasMelcherVenakides}.

\subsection{Small radius}
\label{sec:smallRadius}

\begin{figure}
    \centering
    (a)     \includegraphics[width=8cm]{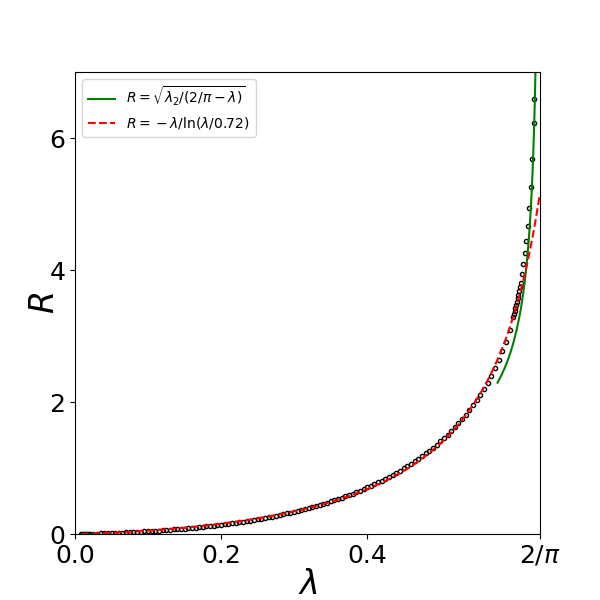}
    
    (b)\includegraphics[width=8cm]{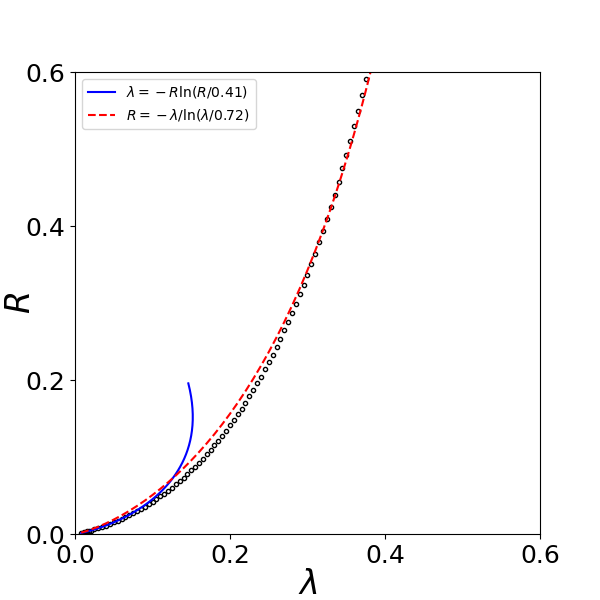}
    \caption{(a) The skyrmion radius $R$, in units of $\ldw$, found numerically for various values of $\dm$, defined in Eq.~\eqref{eq:parameter_aniso}, is shown by open circles.
    The red dashed line shows the fitting formula \eqref{eq:dm-R_aniso_fit}.
    The green solid line shows the asymptotic formula \eqref{eq:dmR_aniso_large_fit} for large radius.
    Figure (b) is a blow up of figure (a).
    The blue solid line shows the asymptotic formula \eqref{eq:dmR_aniso} for small radius.
}
    \label{fig:dm-R_aniso}
\end{figure}

For skyrmions of small radius, it is convenient to work in the variables $u(r)$ of Eq.~\eqref{eq:stereo} and $\v(r)$ of Eq.~\eqref{eq:v} (see Appendix~\ref{sec:asymptotics}).
For $\anisotropy=1$ and $\hext=0$, Eq.~\eqref{eq:dm_R} reduces to
\begin{equation} \label{eq:dm_v0-anisotropy}
    \dm = -R\, [\gamma + 1 + \ln(R/2)]
\end{equation}
or
\begin{equation} \label{eq:dmR_aniso}
  \dm = -R\, \ln\left( \frac{R}{\alpha_\anisotropy} \right),
  \qquad \alpha_\anisotropy = 2 e^{-(\gamma+1)} \approx 0.4131.
\end{equation}
This asymptotic formula was derived in Ref.~\cite{2020_NL_KomineasMelcherVenakides}.
It is shown in Fig.~\ref{fig:dm-R_aniso}a that it matches excellently the numerically found skyrmion radius for small $R$ in the range $0 < \dm \lesssim 0.1$.

Inverting formula \eqref{eq:dmR_aniso} gives, to the lowest order of approximation, Eq.~\eqref{eq:dm-R_small}.
A modification of the denominator of Eq.~\eqref{eq:dm-R_small},
\begin{equation} \label{eq:dm-R_aniso_fit}
    R = - \frac{\dm}{\ln(\dm/0.72)},
\end{equation}
as obtained in Ref.~\cite{2022_ZimmermannBluegel}, leads to a very good fit of the numerical data over the very wide range $0 < \dm \lesssim 0.6$, as shown in Fig.~\ref{fig:dm-R_aniso}a.


Using asymptotic results from Appendix~\ref{sec:asymptotics}, we can give approximation formulae for the skyrmion profile.
The field close to the skyrmion center is given in Eq.~\eqref{eq:innerCoreField}.
In the present case ($\anisotropy=1, \hext=0$), this reduces to a formula identical to Eq.~\eqref{eq:core_theta_small} in Sec.~\ref{sec:externalField}.
The far field is given in Sec.~\ref{sec:farField} and it is identical to Eq.~\eqref{eq:farField_theta_small} in Sec.~\ref{sec:externalField}. 

\begin{figure}
    \centering
    (a) \includegraphics[width=8cm]{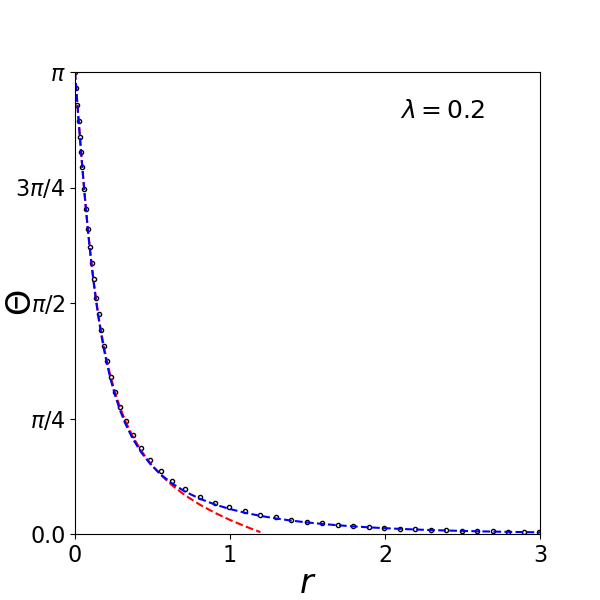}
    
    (b) \includegraphics[width=8cm]{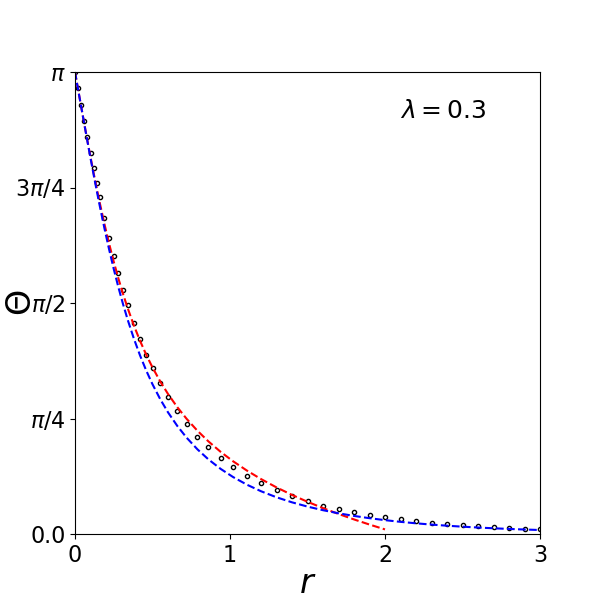}
    \caption{Numerically calculated skyrmion profiles $\Theta(r)$ are shown by open circles for the case of anisotropy and for two values of the parameter (a) $\dm=0.2$ and (b) $\dm = 0.3$ both of which give a small radius.
    The red dashed line shows formula \eqref{eq:core_theta_small} that is valid close to the skyrmion center, and the blue dashed line shows the far field formula \eqref{eq:farField_theta_small}.
    Both formulae turn out to give good approximations far beyond their expected range of asymptotic validity.
    Lengths are measured in units of $\ldw$.}
    \label{fig:profiles_aniso_small}
\end{figure}

Fig.~\ref{fig:profiles_aniso_small} shows formulae \eqref{eq:core_theta_small} and \eqref{eq:farField_theta_small} compared to numerically calculated skyrmion profiles for two values of the parameter.

\subsection{Large radius}

At the parameter value $\dm=2/\pi$, the skyrmion radius diverges to infinity and a phase transition occurs from the uniform to the spiral state.
When the skyrmion is large, an asymptotic series for the profile is given in negative powers of $R$ \cite{2020_NL_KomineasMelcherVenakides},
\begin{equation}\label{eq:Theta_series}
     \Theta = \Theta_0 + \tilde{\Theta},\qquad
     \tilde{\Theta} = \frac{\Theta_1}{R}+\frac{\Theta_2}{R^2}+\frac{\Theta_3}{R^3}+\cdots
\end{equation}
where $\Theta_0, \Theta_1, \Theta_2, \Theta_3,\cdots$ are functions of $r-R$.
The leading order contribution satisfies
\begin{equation} \label{eq:DWprofile}
\tan \left(\frac{\Theta_0}{2}\right) = e^{-(r-R)}.
\end{equation}
This is the standard one-dimensional wall profile.

\begin{figure}
    \centering
    \includegraphics[width=8cm]{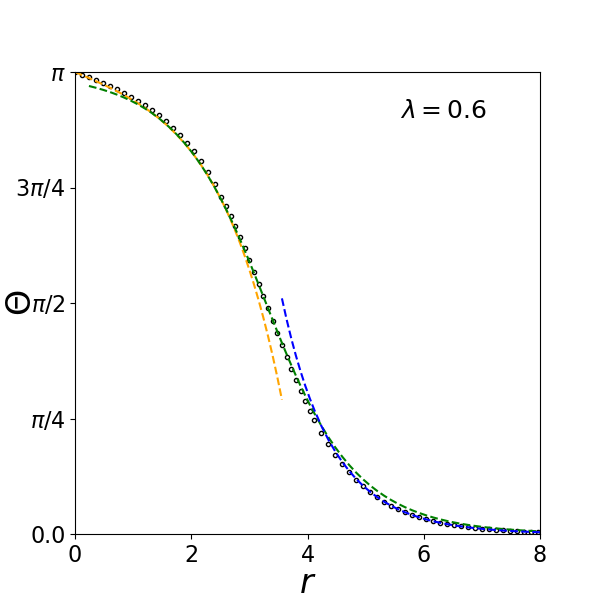}
    \caption{
    Numerically calculated skyrmion profile $\Theta(r)$ shown by open circles for the case of anisotropy and for parameter value $\dm=0.6$ that gives a radius $R=3.29$.
    The orange dashed line shows formula \eqref{eq:Theta_core_large} that is valid in the skyrmion core, the blue dashed line shows the far field formula \eqref{eq:Theta_farField_large}, and the green dashed line shows the field at the domain wall given in Eq.~\eqref{eq:DWprofile}.
    Lengths are measured in units of $\ldw$.}
    \label{fig:profiles_large}
\end{figure}

We now give approximation leading order formulae for the core and the far field of the skyrmion.
The field at the core is \cite{2021_PhysD_KomineasMelcherVenakides}
\begin{equation}  \label{eq:Theta_core_large}
    \Theta(r) = \pi - e^{-R}\sqrt{8\pi R}\, I_1(r)
\end{equation}
where $I_1(r)$ is the modified Bessel function of the first kind.
For the far field, we have
\begin{equation} \label{eq:Theta_farField_large}
    \Theta(r) = e^R \sqrt{\frac{8R}{\pi}}\, K_1(r)
\end{equation}
that simplifies to
\begin{equation} \label{eq:Theta_farField_large_asympt}
    \Theta(r) = 2 \;\frac{e^{-(r-R)}}{\sqrt{r/R}},\qquad r - R \gg 1
\end{equation}
in the region far from the domain wall.
The profile at the skyrmion domain wall region  \eqref{eq:DWprofile} is matched to the skyrmion core and to the far field to all asymptotic orders $O(R^{-n})$ in Ref.~\cite{2021_PhysD_KomineasMelcherVenakides}.

Fig.~\ref{fig:profiles_large} shows  a numerically calculated skyrmion profile and this is compared to the leading order formulae \eqref{eq:DWprofile}, \eqref{eq:Theta_core_large}, and \eqref{eq:Theta_farField_large}.

The relation between the parameter $\dm$ and the radius $R$ is derived during the asymptotic process.
It is given as the series \cite{2020_NL_KomineasMelcherVenakides}, 
\begin{equation}  \label{eq:epsilon_expansion}
\dm = \frac{2}{\pi} + \frac{\dm_2}{R^2} + \frac{\dm_4}{R^4} + \cdots,\qquad R \gg 1
\end{equation}
where
\begin{equation}  \label{eq:epsilon2}
    \dm_2 \approx -0.3057,\quad\dm_4 \approx -0.8792.
\end{equation}

Retaining terms up to $O(1/R^2)$ and solving for $R$, we obtain 
\begin{equation} \label{eq:dmR_aniso_large_fit}
R = \frac{|\dm_2|^{1/2}}{\left(\frac{2}{\pi}-\dm \right)^{1/2}}.
\end{equation}
This is shown in Fig.~\ref{fig:dm-R_aniso} by the green line.
The fitting formula \eqref{eq:dm-R_aniso_fit} and the asymptotic result \eqref{eq:dmR_aniso_large_fit} for large radius give a very good approximation of the numerically found radii for the entire parameter space.

\section{Magnetostatic field}
\label{sec:magnetostatic}

The magnetostatic field is known to be crucial for stabilising magnetic bubbles  \cite{MalozemoffSlonczewski} which are cylindrical domains that share topological features with skyrmions.
The relation between skyrmions and bubbles has been already explored \cite{2018_SciPost_MantelCamosi,2018_SciRep_BuettnerLemeshBeach,2018_SciRep_BuettnerLemeshBeach}.
Given the role of the magnetostatic field in stabilizing magnetic bubbles, we expect that it will often (but not always) have a stabilizing effect also on chiral skyrmions.
When the DM interaction is present, the magnetostatic field is usually considered to be of secondary importance for skyrmion generation and stability.
In this section, we present the results of numerical simulations for the profile of skyrmions when the magnetostatic field is included.
We also give a comparison between bubble and skyrmion profiles.

We first summarize the results on the stability of cylindrical domains \cite{1969_BSTJ_Thiele} that will be useful in the interpretation of the numerical results in this section.
Easy-axis anisotropy perpendicular to the film is necessary in order to have stable magnetic bubbles in a material.
The magnetostatic field is favoring the expansion of the bubble and it thus has a demagnetization effect.
The bubble domain wall tends to shrink in order to minimize the bubble size, however, the demagnetizing effect is typically stronger and thus no energy balance can be achieved.
The stabilization of bubbles is actually obtained by the addition of an external (bias) field perpendicular to the film.

Magnetic bubbles, even ones with high skyrmion numbers, have been observed and studied \cite{MalozemoffSlonczewski}.
Topologically trivial magnetic bubbles, with skyrmion number zero, have also been observed.
The bubble domain wall is primarily of Bloch type as this is favoured by the magnetostatic interaction.
Both Bloch chiralities are energetically equivalent.
The structure of the bubble wall is though complicated by the effects of the film boundaries \cite{1979_JAP_BlakeDellaTorre,KomineasPapanicolaou_PhysD1996}.

The static Landau-Lifshitz equation has the form
\begin{equation} \label{eq:LL-magnetostatic}
\magn \times \left( \p_\mu\p_\mu\magn - 2\dm\, \e_\mu\times\p_\mu\magn + m_3 \ez + \frac{\hext}{\anisotropy}\ez + \frac{\hmagn}{\anisotropy} \right) = 0
\end{equation}
where the bulk DM interaction term has been chosen, $\hmagn$ is the magnetostatic field normalized to the saturation magnetization, and $\ldw$ is used as the unit of length.
The parameters are given here again for convenience
\begin{equation}  \label{eq:parameter_magneto}
\dm = \frac{\DM}{2\sqrt{A\Anisotropy}},\qquad
\anisotropy = \frac{2\Anisotropy}{\mu_0 M_s^2}.
\end{equation}

We often consider the substitution $\Anisotropy \to \Anisotropy - \frac{1}{2}\mu_0 M_s^2$ in order to take into account the fact that the magnetostatic field is equivalent to easy plane anisotropy for ultra-thin films.
This substitution is equivalent to seting $\anisotropy\to\anisotropy-1$ and $\dm \to \anisotropy/(\anisotropy-1)\dm$ in Eq.~\eqref{eq:LL-magnetostatic}.
After these substitutions, the field $\hmagn$ would represent only the deviation of the magnetostatic field from its approximation as an easy-plane anisotropy term. 

We find solutions of Eq.~\eqref{eq:LL-magnetostatic} using an energy relaxation algorithm \cite{KomineasPapanicolaou_PhysD1996}.
Our method works in cylindrical coordinates $(r,z)$ and it is confined to axially symmetric configurations only.
The magnetostatic field is calculated using a conjugate gradient method.
We consider an infinite film achieved by employing a long enough numerical mesh in the $r$ direction and assuming no magnetic charges on the lateral boundary.
We place the film of thickness $\thickness$ so that its central plane is at $z=0$, {\it i.e.}, the film extends over $-\thickness/2 \leq z \leq \thickness/2$.
We find that solutions for the magnetization $\magn=(m_r,\ m_\phi, m_z)$ satisfy the parity relations $m_r(r,-z)=-m_r(r,z),\, m_\phi(r,-z)=m_\phi(r,z),\, m_z(r,-z)=m_z(r,z)$.

\begin{figure}[t]
    \centering
    (a) \includegraphics[width=0.85\columnwidth]{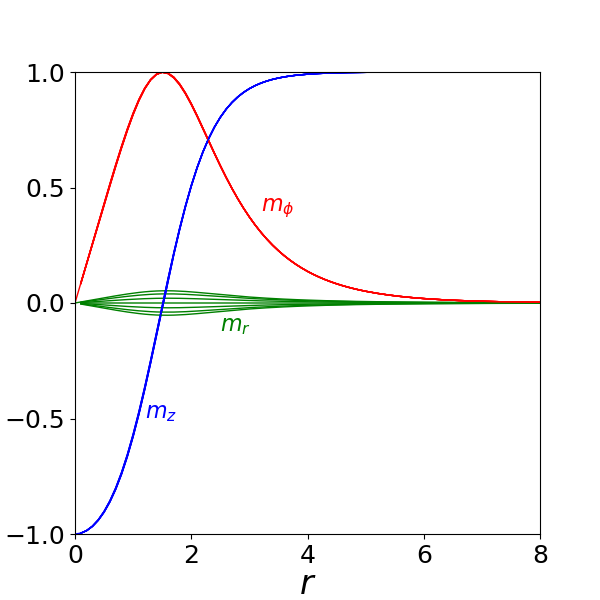}
    
    (b) \includegraphics[width=0.85\columnwidth]{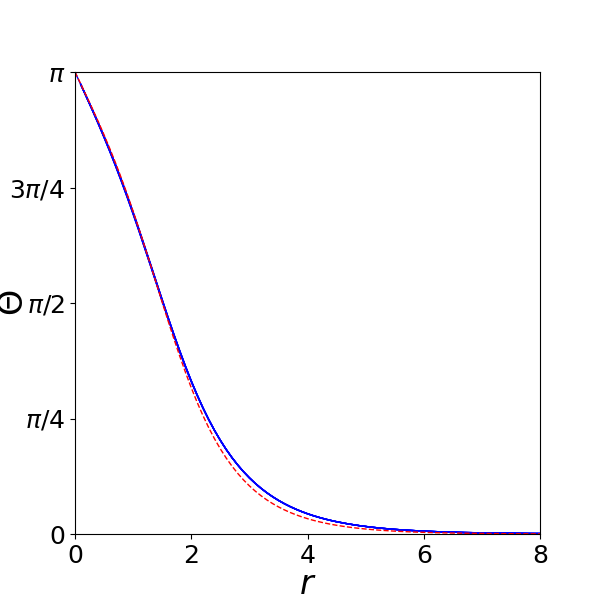}
    \caption{(a) The cylindrical components of the magnetization vector $\magn=(m_r, m_\phi, m_z)$ for a chiral skyrmion including the magnetostatic interaction in a film of thickness $\thickness=3.5$ (the film extends over $-1.75 \leq z \leq 1.75)$, for parameter values $\dm=0.33,\; \anisotropy=4$.
    We show the magnetization as function of the radial coordinate $r$ at the levels $z=0,\pm 0.5, \pm 1.0, \pm 1.5$.
    It is $m_r > 0$ for $z<0$ and $m_r < 0$ for $z>0$.
    The components $m_\phi, m_z$ depend very weakly on $z$ (it is barely visible in the figure).
    (b) The blue line shows the profile $\Theta(r)$ of the skyrmion.
    The red dashed line shows the profile of a skyrmion with a similar radius found without the magnetostatic interaction, obtained for $\dm=0.51$.
    Lengths are measured in units of $\ldw$.
    }
    \label{fig:thinFilm_magnetization}
\end{figure}

We first explore the case of easy axis anisotropy only and no bias field.
We consider a thin film with thickness $\thickness=3.5\ldw$ and parameter values $\dm=0.33,\;\anisotropy=4$.
We discretize space with lattice spacings $\Delta r = 0.1$ and $\Delta z = 0.5$ in the $r$ and $z$ directions respectively.
The numerical mesh extends to $r=20$.
Fig.~\ref{fig:thinFilm_magnetization} shows the cylindrical components of the magnetization for a skyrmion solution as functions of $r$ at various levels of $z$.
The skyrmion radius is $R\approx 1.5$, a significant increase in comparison to the skyrmion radius $R=0.42$ found for $\dm=0.33$ when the magnetostatic field is neglected.
Furthermore, the profile acquires a nonzero $m_r$ component, which is an odd function in the direction perpendicular to the film.

When the parameter $\dm$ is increased, the skyrmion radius increases.
In the presence of the magnetostatic field, we find no static skyrmions for $\dm > 0.35$.
This should be compared to the critical value $\dm=2/\pi$ when the magnetostatic field is neglected.
This finding suggests that skyrmions with a large radius cannot be obtained (without a bias field) because they are destabilised by the magnetostatic field.

\begin{figure}[t]
    \centering
(a) \includegraphics[width=0.85\columnwidth]{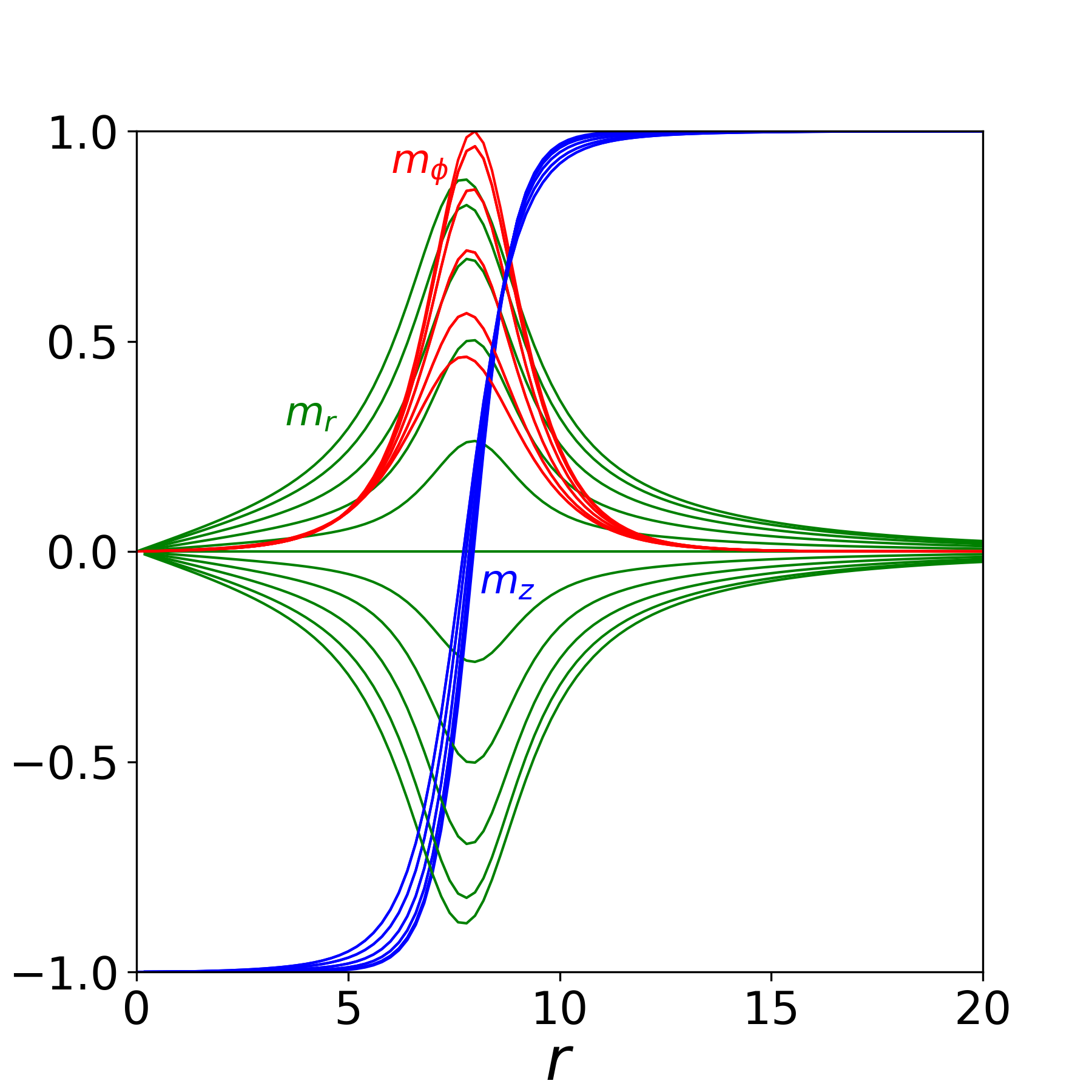}

(b) \includegraphics[width=0.85\columnwidth]{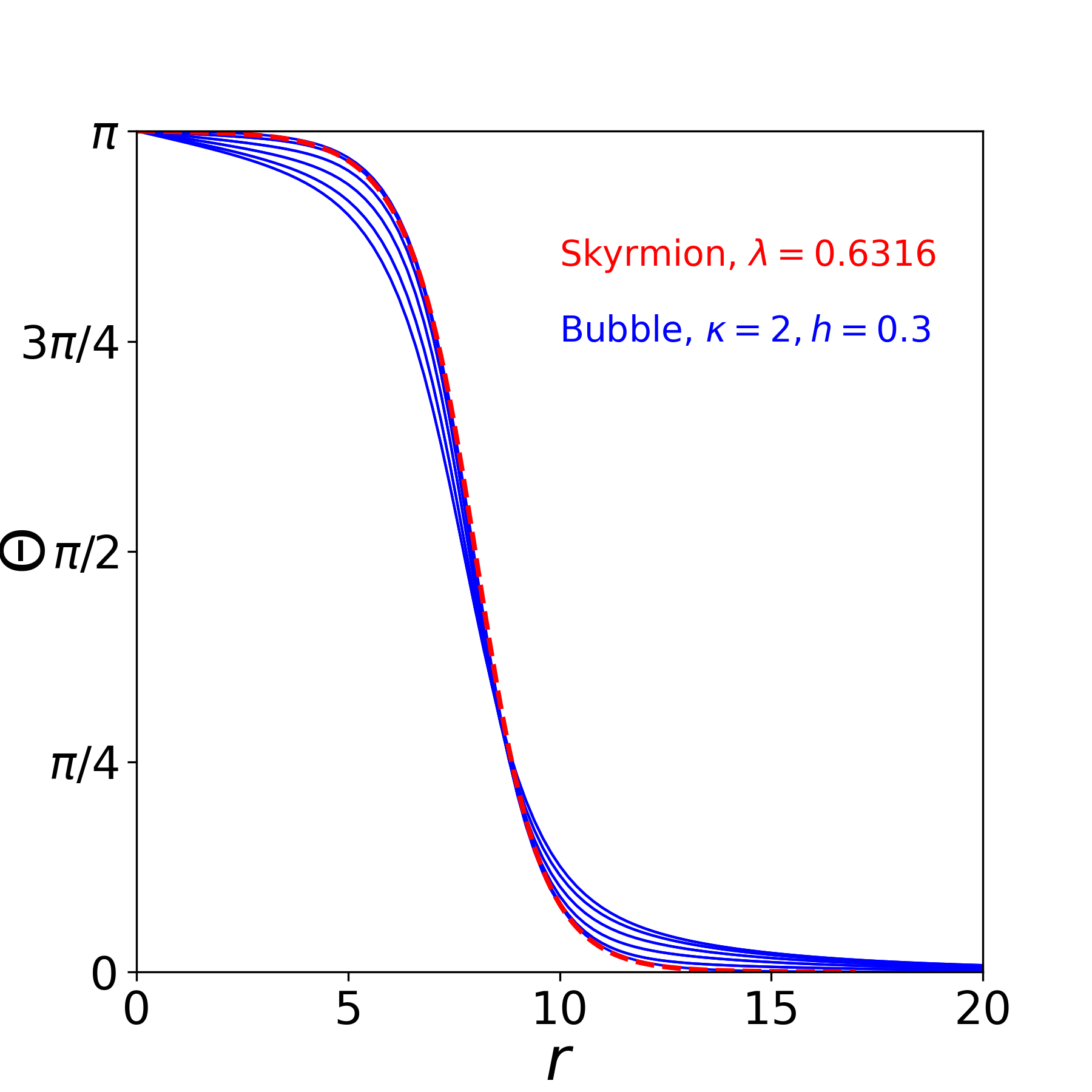}
    \caption{
    (a) The cylindrical components of the magnetization vector $\magn=(m_r, m_\phi, m_z)$ for a magnetic bubble in a film of thickness $\thickness=11$ (the film extends over $-5.5 \leq z \leq 5.5$), for parameter values $\anisotropy=2,\;\hext=0.3$ (no DM interaction, $\dm=0$).
    We show the magnetization as function of the radial coordinate $r$ at the levels $z=0,\pm1, \pm2, \pm3, \pm4, \pm5$.
    It is $m_r > 0$ for $z<0$ and $m_r < 0$ for $z>0$.
    (b) The blue lines show the profile $\Theta(r)$ of the bubble at the same $z$-levels.
    The red dashed line shows the profile of a chiral skyrmion with a radius similar to that of the bubble.
    It is obtained for $\dm=0.6316$ when the magnetostatic field is neglected.
    Lengths are measured in units of $\ldw$.}
    \label{fig:profile_bubble_magn}
\end{figure}

It is interesting to compare the chiral skyrmion profiles with magnetic bubble ones that are stabilised without the DM interaction.
We choose a film with thickness $\thickness=11\ldw$, extending over $-\thickness/2 \leq z \leq \thickness/2$, and the parameter values $\anisotropy=2,\;\hext=0.2$ (and set $\dm=0$).
The lattice spacings are $\Delta r = 0.2,\;\Delta z = 1.0$ and the numerical mesh extends to $r=40$.
The energy minimization algorithm converges to a static bubble with radius $R\approx 8$.
Fig.~\ref{fig:profile_bubble_magn}a shows the three components of the magnetization as functions of $r$ at various levels of $z$.
The stray field at the bubble domain wall induces large values for $m_r$ close to the film boundaries.
Correspondingly, the component $m_\phi$ depends strongly on $z$.
As a result, the bubble wall is of a hybrid character; Bloch at the center and tilted towards Ne\'el type closer to the boundaries.
The hybrid wall structure has been shown to persist when DM interaction is included \cite{2018_SciAdv_LegrandCrosFert,2019_AdvMater_LiBykovaZhang}.
The component $m_z$ also has a dependence on $z$ that results in the bubble domain wall width to depend on $z$ and to the bulging shape of the bubble \cite{1979_JAP_BlakeDellaTorre,KomineasPapanicolaou_PhysD1996}.

Fig.~\ref{fig:profile_bubble_magn}b shows the spherical angle $\Theta$ for the bubble by blue solid lines at various $z$-levels.
For comparison, a red dashed line shows the chiral skyrmion profile that is obtained for a DM parameter $\dm=0.6316$ and no magnetostatic field.
For that value of $\dm$, the chiral skyrmion has a similar radius as the bubble in the figure.
We finally note that our comparison between a chiral skyrmion and a bubble is only possible for large radii as magnetic bubbles are not stabilised for small radii \cite{1969_BSTJ_Thiele}.

\section{Energy of a skyrmion}
\label{sec:energy}

The skyrmion radius, studied in the previous sections, is a length scale that determines the balance of the individual interaction terms and thus provides information about the skyrmion energy.
The energetics of the model is important not only as a means to understand the skyrmions as energy minima but also in order to build arguments concerning their stability and excitations.
In the regime of small radii, the energy asymptotics has been examined rigorously in \cite{2017_DoeringMelcher, 2020_ARMA_BernandMuratovSimon,2022_GustafsonWang}.

We consider the case of easy-axis anisotropy and no external field so that the energy is given in Eq.~\eqref{eq:energy} for $\anisotropy=1, \hext=0$,
and it is measured in units of $\mu_0 M_s^2 \lex^2 = 2A$.
A useful result for the energy components of a skyrmion is obtained by employing a standard scaling argument \cite{BogdanovHubert_JMMM1994},
\begin{equation} \label{eq:virial}
    2\Ean + \Edm = 0.
\end{equation}

\subsection{Small radius}

For small skyrmion radius, asymptotic analysis gives \cite{2020_NL_KomineasMelcherVenakides}
\begin{equation} \label{eq:energy_small_asymptotic}
    \Eex = 4\pi + o\left(\frac{\dm^2}{\ln\dm} \right), \quad \Edm = 8\pi \frac{\dm^2}{\ln\dm} + o\left(\frac{\dm^2}{\ln\dm} \right)
\end{equation}
while $\Ean$ can be obtained from Eq.~\eqref{eq:virial}.
Result \eqref{eq:energy_small_asymptotic} can be used to produce asymptotic series for the energy terms as we will now demonstrate.

We consider a specific DM parameter $\dm_0$ and the energy $\Energy=\Energy_{\dm_0}(\Theta)$ for any configuration $\Theta$.
This is written as
\begin{equation} \label{eq:energy_dm0}
\Energy_{\dm_0}(\Theta) = \Eex(\Theta) + \dm_0 \Edmm(\Theta) + \Ean(\Theta)
\end{equation}
where we have set
\begin{equation} \label{eq:Edmm}
\Edm = \dm_0 \Edmm,
\end{equation}
so that all terms $\Eex, \Edmm, \Ean$ are integrals over $\Theta$ that do not contain parameters.

We confine ourselves to the specific configurations $\Theta=\Theta_\dm$ that are the profiles corresponding to static skyrmions for any parameter value $\dm$ (that may be different from $\dm_0$).
The virial relation \eqref{eq:virial} holds,
\begin{equation}
\Ean = -\frac{\dm}{2}\Edmm,
\end{equation}
and it is used in Eq.~\eqref{eq:energy_dm0} to obtain
\begin{equation} \label{eq:EexEdmm}
\Energy_{\dm_0}(\dm) = \Eex + \left(\dm_0-\frac{\dm}{2} \right)\Edmm
\end{equation}
for the energy of profile $\Theta_\dm$ in a system with parameter value $\dm_0$.

We take the derivative of the energy expression \eqref{eq:EexEdmm} with respect to $\dm$,
\begin{equation}
\Energy_{\dm_0}' = \Eex' + \left( \dm_0 - \frac{\dm}{2} \right) \Edmm' - \frac{1}{2}\Edmm
\end{equation}
and require that $\Energy_\dm'(\dm=\dm_0)=0$.
This gives the condition
\begin{equation} \label{eq:conditionForMinimumAtdm0}
\Eex' + \frac{1}{2} \left(\dm \Edmm' - \Edmm \right) = 0\qquad \text{at}\; \dm=\dm_0.
\end{equation}
An explicit calculation of the terms of Eq.~\eqref{eq:conditionForMinimumAtdm0} utilizing Eq.~\eqref{eq:energy_small_asymptotic}, shows that terms of order $O(\dm^2/(\ln\dm)^m)$ with $m\ge 2$ should necessarily be included in the energy asymptotic series for the condition \eqref{eq:conditionForMinimumAtdm0} to be satisfied.
Specifically,  
\begin{equation} \label{eq:EexEdmm_series}
\begin{split}
\Eex & = 4\pi + 4\pi \sum_{m=2}^\infty a_m\,\frac{\dm^2}{(\ln\dm)^m},\\ \Edmm & = 4\pi \sum_{m=1}^\infty b_m\,\frac{\dm}{(\ln\dm)^m}
\end{split}
\end{equation}
where $a_m, b_m$ are constants (with $b_1=2$ from Eq.~\eqref{eq:energy_small_asymptotic}).
Substituting these in Eq.~\eqref{eq:conditionForMinimumAtdm0} obtains
\begin{equation} \label{eq:satisfySeries}
\sum_{m=1}^\infty \frac{2 a_{m+1} - m a_m - \frac{m}{2} b_m)}{(\ln\dm)^{m+1}} = 0,\quad m=1,2,\cdots.
\end{equation}
In order that Eq.~\eqref{eq:satisfySeries} be satisfied to all orders in $1/(\ln(\dm))^m$, it should hold
\begin{equation} \label{eq:a_m1}
     a_{m+1} = \frac{m}{4} \left( 2a_m + b_m \right)\quad m=1,2,\cdots.
\end{equation}
For $m=1$, this gives (since $a_1=0$)
\begin{equation}
a_2 = \frac{b_1}{4} = \frac{1}{2}.
\end{equation}
We thus have
\begin{equation} \label{eq:Eex_asymptotic2}
\Eex = 4\pi \left( 1 + \frac{1}{2}\,\frac{\dm^2}{(\ln\dm)^2} \right) + O\left( \frac{\dm^2}{(\ln\dm)^3} \right).
\end{equation}

More information would be required to obtain explicit values for $a_{m+1}, b_m$ with $m\geq 2$.
Comparing the second of Eqs.~\eqref{eq:EexEdmm_series} with numerical data, we propose the following formula,
\begin{equation} \label{eq:Edm_asymptotic2}
    \Edm = 4\pi \left( 2\frac{\dm^2}{\ln\dm} - \frac{1}{2} \frac{\dm^2}{(\ln\dm)^2} \right) + O\left( \frac{\dm^2}{(\ln\dm)^3} \right).
\end{equation}
Using formulae \eqref{eq:Eex_asymptotic2} and \eqref{eq:Edm_asymptotic2}, the energy is
\begin{equation} \label{eq:E-dm_small}
    E = 4\pi \left( 1 + \frac{\dm^2}{\ln\dm} + \frac{1}{4} \frac{\dm^2}{(\ln\dm)^2} \right) + O\left( \frac{\dm^2}{(\ln\dm)^3} \right).
\end{equation}
This is in agreement with rigorous results in Ref.~\cite{2022_GustafsonWang} where an error term $O(\dm^2/(\ln\dm)^2)$ is given.
The numerical data for the energy are fitted well with formulae \eqref{eq:Eex_asymptotic2}, \eqref{eq:Edm_asymptotic2}, \eqref{eq:E-dm_small}, for parameter values $\dm \lesssim 0.25$.

\begin{figure}
    \centering
    \includegraphics[width=8cm]{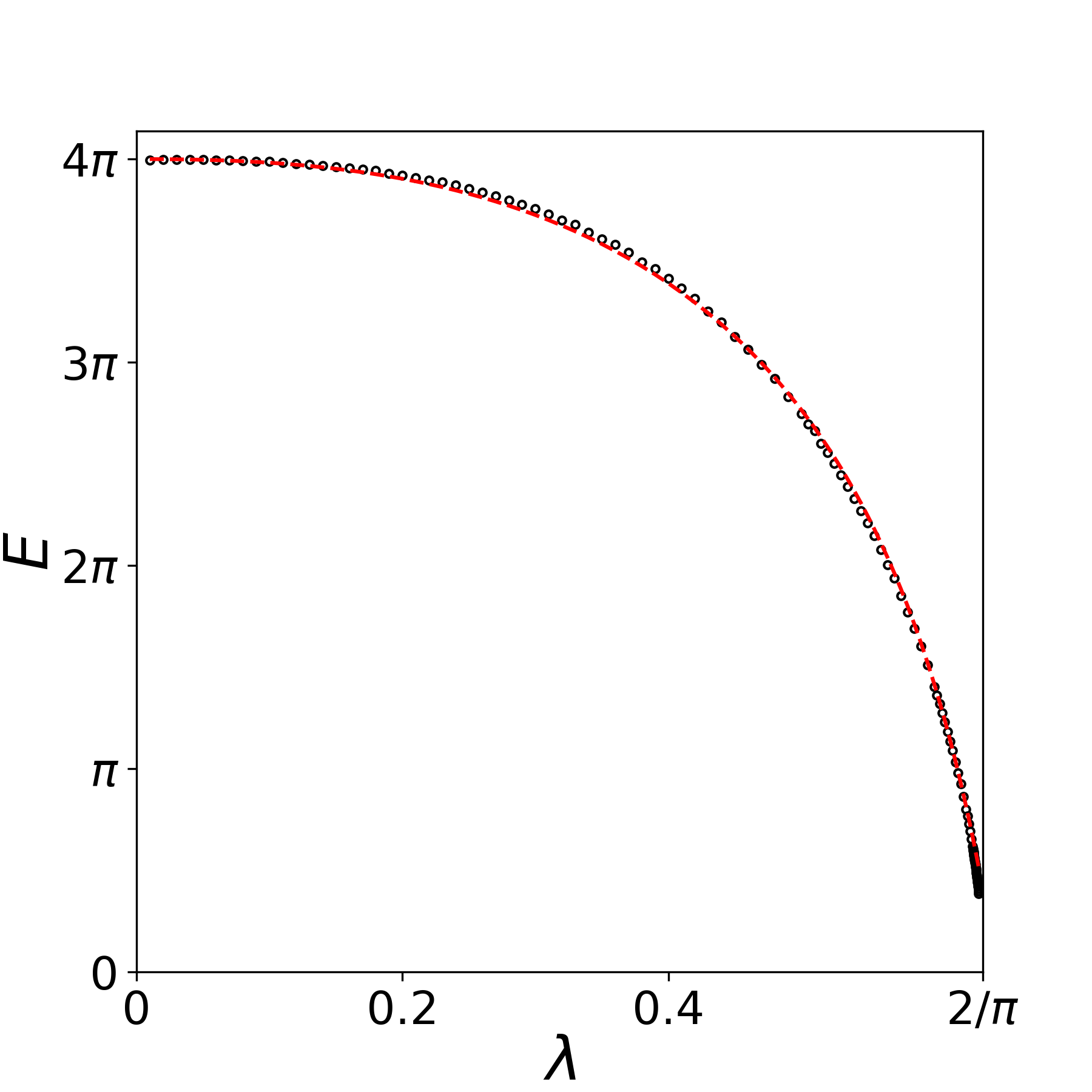}
    \caption{The skyrmion energy as a function of the parameter $\dm$ for the case of the model with anisotropy.
    The open circles show numerically calculated data.
    The dashed red line shows formula \eqref{eq:E-dm_small_adjusted}.}
    \label{fig:E-dm}
\end{figure}

Fig.~\ref{fig:E-dm} shows the numerically calculated skyrmion energy as a function of $\dm$.
A modification of formula \eqref{eq:E-dm_small} is found to fit the numerical data very well almost in the entire parameter range,
\begin{equation} \label{eq:E-dm_small_adjusted}
    E = 4\pi \left( 1 + \frac{\dm^2}{\ln(\dm/0.75)} (1-\dm) \right).
\end{equation}
This formula is plotted in Fig.~\ref{fig:E-dm}.
For $\dm$ close to the critical value $2/\pi$ formula \eqref{eq:E-dm_large} (of the next subsection) should be used instead.

Eq.~\eqref{eq:E-dm_small} gives
\begin{equation} \label{eq:E-R_small}
\begin{split}
    E \approx 4\pi \left( 1 + R^2\ln R \right),\quad R \ll 1.
\end{split}
\end{equation}
Eq.~\eqref{eq:E-R_small} gives a good approximation only in the range $R < 0.5$.
It is remarkable that the numerical data are fitted well over a large range of $R$ by a quite significant modification of Eq.~\eqref{eq:E-R_small},
\begin{equation} \label{eq:E-R_small_modified}
    E = 4\pi \left[ 1 + 0.12\,R^2\ln(R/5.2) \right],
\end{equation}
that is plotted in Fig.~\ref{fig:energy-rad_anisotropy}.

We finally note that the techniques used in this section can be readily applied to the case of applied external field.

\subsection{Large radius}

For large skyrmion radius, asymptotic analysis gives \cite{2021_PhysD_KomineasMelcherVenakides}
\begin{equation} \label{eq:E-dm_large}
    \Energy \sim 4\pi^2 |\dm|^{1/2}\,\sqrt{\frac{2}{\pi}-\dm},\quad \left|\frac{2}{\pi}-\dm\right| \ll 1
\end{equation}
or (cf. Eq. \eqref{eq:dmR_aniso_large_fit})
\begin{equation} \label{eq:E-R_large}
    \Energy = \frac{4\pi^2|\dm_2|}{R} +O\left( \frac{1}{R^3} \right),\quad R \gg 1.
\end{equation}
In the range $0.6 \lesssim \dm < 2/\pi$, the numerical data of Fig.~\ref{fig:E-dm} are fitted well with formula \eqref{eq:E-dm_large}.

\begin{figure}[t]
    \centering
    \includegraphics[width=8cm]{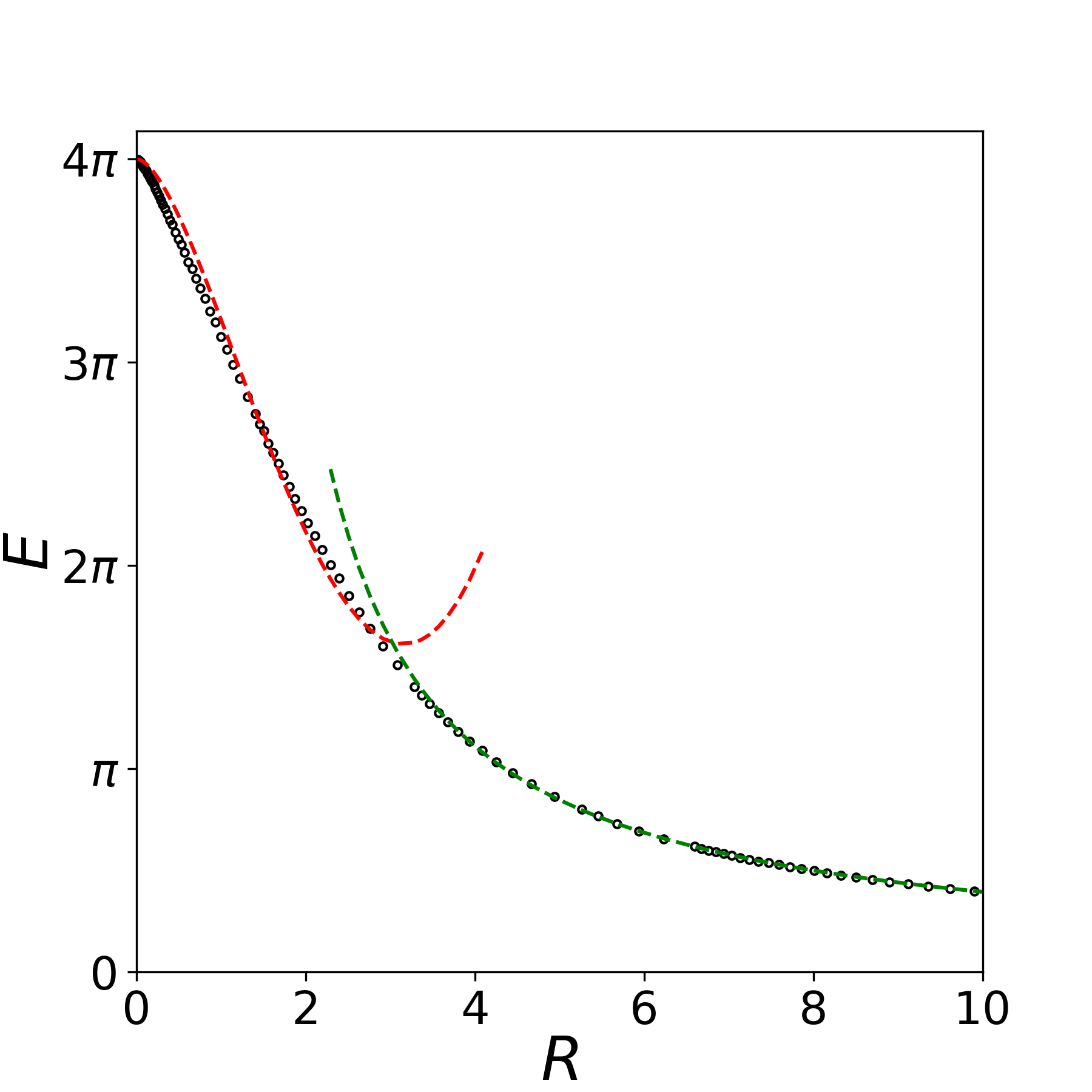}
    \caption{
    The skyrmion energy as a function of the radius $R$ for the case of the model with anisotropy.
    The open circles show numerically calculated data.
    The dashed red line shows formula \eqref{eq:E-R_small_modified} and the dashed green line shows formula \eqref{eq:E-R_large_adjusted}.}
    \label{fig:energy-rad_anisotropy}
\end{figure}

Fig.~\ref{fig:energy-rad_anisotropy} shows the skyrmion energy as a function of its radius.
For large $R$, the data are fitted very well with the formula
\begin{equation} \label{eq:E-R_large_adjusted}
    E = 4\pi^2|\dm_2| \left( \frac{1}{R} + \frac{5}{2}\frac{1}{R^3} \right)
\end{equation}
which is a sharpening of Eq.~\eqref{eq:E-R_large} and it is plotted in the figure.
We further note that each one of the energy components is fitted very well, for $R \gtrsim 3$, by the forms
\begin{equation}
\begin{split}
    \Eex & = 2\pi R + \pi^2|\dm_2| \left(\frac{3}{R} + \frac{5}{R^3} \right) \\
    \Ean & = 2\pi R - \pi^2|\dm_2| \left(\frac{1}{R} + \frac{5}{R^3} \right).
\end{split}
\end{equation}

\section{Concluding remarks}
\label{sec:conclusions}

We have presented fitting formulae of skyrmion profiles and energy that cover almost the entire range of the parameter values and of skyrmion radii.
These are based on formulae obtained through asymptotic analysis \cite{2020_NL_KomineasMelcherVenakides,2021_PhysD_KomineasMelcherVenakides}.
The analysis is extended to the case of an external magnetic field in the Appendix.
We were thus able to study both mechanisms that give stable skyrmions, i.e., the case of an external field and the case of easy-axis anisotropy, and we have given asymptotic formulae when both terms are present in the model.

We have further studied the effect of the magnetostatic field on the skyrmion profile.
It tends to increase the skyrmion radius while it appears to destabilise skyrmions of large radius.
We compared the profile of magnetic bubbles stabilized without the chiral Dzyaloshinskii-Moriya interaction to that of a chiral skyrmion.
We found that the profiles are similar in the two cases, the main difference between the two being the variation of the bubble profile across the film thickness. 

Finally, we have produced asymptotic series expressions for all skyrmion energy terms using an argument for the minimization of the energy.
This has resulted in numerical expressions that sharpen previous asymptotic results.
We also give approximate formulae that cover almost the entire range of the parameter values and of skyrmion radii.

The obtained formulae can be used for comparisons with experimental results and for a variety of investigations where the details of skyrmion features are important, notably, in skyrmion dynamics.

\bigskip
\bigskip

\onecolumngrid

\appendix

\section{Asymptotic analysis for skyrmion of small radius}
\label{sec:asymptotics}

\subsection{Near field}
\label{sec:nearField}

We will mainly work with the field $\v(r)=ru(r)$ defined in Eq.~\eqref{eq:v}.
It satisfies the equation
\begin{equation}  \label{eq:v-prime}
\v'' + \frac{\v'}{r}\, \frac{3\v^2-r^2}{\v^2+r^2} - \frac{2 \v \v'^2}{\v^2 + r^2} + 4\dm \frac{\v^2}{\v^2+r^2} + \anisotropy\frac{\v^2-r^2}{\v^2+r^2} \v - \hext \v = 0.
\end{equation}
The ensuing analysis will follow the techniques developed in Ref.~\cite{2020_NL_KomineasMelcherVenakides}.

An assumption which is found to be consistent with our results is to neglect the term containing $\v'^2$ in Eq.~\eqref{eq:v-prime},
\begin{equation}  \label{eq:v-prime_nearField0}
\v'' + \frac{\v'}{r}\, \frac{3\v^2-r^2}{\v^2+r^2} + 4\dm \frac{\v^2}{\v^2+r^2} + \anisotropy\frac{\v^2-r^2}{\v^2+r^2} \v - \hext \v = 0.
\end{equation}
In the near field, we replace $\v$ in Eq.~\eqref{eq:v-prime_nearField0} with its initial value $\v(0)=\v_0$.
The approximation is valid for the range of $r$ over which $\v_0-\v\ll \v_0$. 
The equation obtained in this way,
\begin{equation}  \label{eq:vs_nearField}
\v'' + \frac{\v'}{r}\, \frac{3\v_0^2-r^2}{\v_0^2+r^2} = -4\dm \frac{\v_0^2}{\v_0^2+r^2} - \anisotropy\frac{\v_0^2-r^2}{\v_0^2+r^2} \v_0 + \hext \v_0
\end{equation}
is integrable by the integrating factor
\[
\mu(r) = \frac{r^3}{(\v_0^2+r^2)^2}.
\]
We obtain
\[
\frac{d}{dr} \left[ \frac{r^3}{(\v_0^2+r^2)^2} v' \right] = -4\dm \frac{\v_0^2 r^3}{(\v_0^2+r^2)^3}
  -\anisotropy\frac{\v_0^2 - r^2}{(\v_0^2+r^2)^3} \v_0 r^3 + \hext \frac{\v_0 r^3}{(\v_0^2+r^2)^2},
\]
which integrates to
\[
\frac{r^4}{(\v_0^2+r^2)^2}\frac{\v'}{r} = -\dm \frac{r^4}{(\v_0^2+r^2)^2}
 + \frac{\v_0}{2} \left[ (\anisotropy+\hext) \ln\left( 1 + \frac{r^2}{\v_0^2} \right) - \anisotropy \frac{\v_0^2+2r^2}{(\v_0^2+r^2)^2} r^2 - \hext \frac{r^2}{\v_0^2+r^2} \right].
\]
We finally have
\begin{equation}  \label{eq:nearField_vdot}
\frac{\v'}{r} = -\dm
 + \frac{\v_0}{2}
\left[ (\anisotropy+\hext) \frac{(\v_0^2+r^2)^2}{r^4} \ln\left( 1 + \frac{r^2}{\v_0^2} \right)
- \anisotropy \frac{\v_0^2+2r^2}{r^2} - \hext \frac{\v_0^2+r^2}{r^2}\right].
\end{equation}
The constant of integration has been judiciously chosen to eliminate the fourth-order singularity $r^{-4}$ on the right.

We make the change of variables $w(\tau)=\v/\v_0$ and $\tau=r^2/\v_0^2$.
Then Eq.~\eqref{eq:nearField_vdot} becomes
\begin{equation}  \label{eq:nearField_wdot}
\frac{dw}{d\tau} = -\dm \frac{\v_0}{2}
+(\anisotropy + \hext) \frac{\v_0^2}{4} \left[ \ln(1+\tau) + 2 \frac{\ln(1+\tau)}{\tau} + \frac{\ln(1+\tau)}{\tau^2} - \frac{1}{\tau}  \right] - (2\anisotropy+\hext) \frac{\v_0^2}{4}.
\end{equation}
This is integrated to give
\[
w(\tau) = 1 - \dm\frac{\v_0}{2} \tau
 + (\anisotropy + \hext) \frac{\v_0^2}{4} \left[ \tau \ln(1+\tau) - \frac{\ln(1+\tau)}{\tau} + 1 - 2{\rm Li}_2(-\tau) - \tau \right] - (2\anisotropy+\hext) \frac{\v_0^2}{4} \tau
\]
or
\begin{equation}  \label{eq:nearField_wtau}
w(\tau) = 1 - \left( \dm + \frac{3}{2}\anisotropy \v_0 + \hext \v_0 \right) \frac{\v_0}{2} \tau
 + (\anisotropy + \hext) \frac{\v_0^2}{4} \left[ \tau \ln(1+\tau) - \frac{\ln(1+\tau)}{\tau} + 1 - 2{\rm Li}_2(-\tau) \right]
\end{equation}
where ${\rm Li}_2(x)$ is the dilogarithm function and the constant of integration has been chosen so that $w(0)=1$.

Keeping the dominant terms for large $\tau$ we have
\begin{equation} \label{eq:nearField_wtau_dominantTerms}
w(\tau) = 1 - \left(\dm + \frac{3}{2}\anisotropy \v_0 + \hext \v_0 \right) \frac{\v_0}{2}\tau + (\anisotropy + \hext) \frac{\v_0^2}{4} \tau \ln\tau
\end{equation}
which, in the original variables, gives the near field
\begin{equation} \label{eq:nearField_v_matching}
\v_N(r) = \v_0 - \left[ \dm + \anisotropy \v_0 \left(\frac{3}{2} + \ln \v_0 \right) + \hext \v_0 ( 1 + \ln\v_0 ) \right] \frac{r^2}{2} + (\anisotropy + \hext) \v_0 \frac{r^2}{2} \ln r.
\end{equation}

The skyrmion radius is found by the condition $\v(r=R)=R$.
In the case that the skyrmion radius $R$ is small, this is given by
\begin{equation} \label{eq:R=v0}
    R\approx \v_0.
\end{equation}

\subsection{Asymptotic matching}
\label{sec:matching}

The solution of the equation \eqref{eq:modifiedBessel_theta} for the far field is
\begin{equation} 
\Theta \propto \frac{1}{\s} + (2\gamma-1) \s + 2 \sum_{n=1}^\infty \frac{(\ln \s - \ln n) \s^{2n-1}}{(n-1)!\, n!}
  - 2 \sum_{n=2}^\infty \frac{\xi_n \s^{2n-1}}{(n-1)!\, n!},\qquad \s=\sqrt{\anisotropy+\hext}\,\frac{r}{2}
\end{equation}
where $\xi_n$ is given in Eq.~\eqref{eq:Euler} and the solution is valid for any $\anisotropy, \hext$.
For small values of the radial coordinate, we keep only the first three terms in the series,
\[
\Theta \propto \frac{1}{\s} \left[ 1 + (2\gamma-1) \s^2 + 2 \s^2 \ln \s \right].
\]
The far field for the field $\v$ is derived from the latter formula noting that $\v \approx \frac{r}{2}\Theta$ for small values of $\v, \Theta$, as seen from Eqs.~\eqref{eq:v}, \eqref{eq:stereo}.
We find
\[
\v_F(r) = C \left[ 1 + (\anisotropy + \hext) (2\gamma-1) \frac{r^2}{4} + (\anisotropy + \hext) \frac{r^2}{2} \ln \left(\frac{\sqrt{\anisotropy+\hext}}{2}\,r \right) \right]
\]
or
\begin{equation} \label{eq:farField_v_matching}
\v_F(r) = C \left[ 1 + (\anisotropy + \hext) \left( \gamma-\frac{1}{2} + \ln \left(\frac{\sqrt{\anisotropy+\hext}}{2} \right) \right) \frac{r^2}{2} + (\anisotropy + \hext) \frac{r^2}{2} \ln r \right].
\end{equation}

Matching the near field in Eq.~\eqref{eq:nearField_v_matching} with the far field in Eq.~\eqref{eq:farField_v_matching} order by order we have $C=\v_0$ and
\[
\dm + \anisotropy\v_0 \left( \frac{3}{2} + \ln\v_0 \right) + \hext\v_0 (1+\ln\v_0) = - (\anisotropy + \hext) \left[ \gamma-\frac{1}{2} + \ln \left(\frac{\sqrt{\anisotropy+\hext}}{2} \right) \right]\,\v_0
\]
or
\begin{equation} \label{eq:dm_v0}
  \dm = -\v_0 \left[ \anisotropy (\gamma + 1) + \hext \left( \gamma + \frac{1}{2} \right) + (\anisotropy+\hext) \ln\left( \frac{ \sqrt{\anisotropy+\hext}}{2}\, \v_0 \right) \right].
\end{equation}
Combined with Eq.~\eqref{eq:R=v0}, relation \eqref{eq:dm_v0} gives implicitly the skyrmion radius as a function of the parameters,
\begin{equation} \label{eq:dm_R}
  \dm = -R \left[ \anisotropy (\gamma + 1) + \hext \left( \gamma + \frac{1}{2} \right) + (\anisotropy+\hext) \ln\left( \frac{ \sqrt{\anisotropy+\hext}}{2}\, R \right) \right].
\end{equation}

\subsection{Field at the skyrmion center}
\label{sec:innerCore}

The expression in the square bracket in Eq.~\eqref{eq:nearField_wtau} has the expansion
\[
\tau \ln(1+\tau) - \frac{\ln(1+\tau)}{\tau} + 1 - 2{\rm Li}_2(-\tau) = \frac{5}{2} \tau + \frac{\tau^2}{6} + \ldots,\qquad \tau \ll 1.
\]
Thus, the near field, close to the skyrmion center is
\[
w(\tau) \approx 1 - \left( \dm + \frac{3}{2}\anisotropy \v_0 + \hext \v_0 \right) \frac{\v_0}{2} \tau
 + (\anisotropy + \hext) \frac{\v_0^2}{4} \left( \frac{5}{2} \tau + \frac{\tau^2}{6} \right).
\]
Using Eq.~\eqref{eq:nearField_wtau}, this gives
\begin{equation}
   w(\tau) \approx 1 + (\anisotropy+\hext) \left[ \gamma + \frac{3}{4}+ \ln\left( \frac{ \sqrt{\anisotropy+\hext}}{2}\, \v_0 \right) \right] \frac{\v_0^2}{2}\tau + (\anisotropy+\hext) \frac{\v_0^2}{24}\tau^2. 
\end{equation}
For the field $\v$, the latter gives
\begin{equation} \label{eq:innerCoreField}
    \frac{\v(r)}{\v_0} \approx 1 + (\anisotropy+\hext) \left[ \gamma + \frac{3}{4}+ \ln\left( \frac{ \sqrt{\anisotropy+\hext}}{2}\, \v_0 \right) \right] \frac{r^2}{2} + (\anisotropy+\hext) \frac{r^4}{24\v_0^2},\qquad r \ll \v_0.
\end{equation}

\bigskip

\end{document}